# Subspace Shapes: Enhancing High-Dimensional Subspace Structures via Ambient Occlusion Shading

Bing Wang and Klaus Mueller, *Senior Member, IEEE*

**Abstract**—We test the hypothesis whether transforming a data matrix into a 3D shaded surface or even a volumetric display can be more appealing to humans than a scatterplot since it makes direct use of the innate 3D scene understanding capabilities of the human visual system. We also test whether 3D shaded displays can add a significant amount of information to the visualization of high-dimensional data, especially when enhanced with proper tools to navigate the various 3D subspaces. Our experiments suggest that mainstream users prefer shaded displays over scatterplots for visual cluster analysis tasks after receiving training for both. Our experiments also provide evidence that 3D displays can better communicate spatial relationships, size, and shape of clusters.

## 1 INTRODUCTION

IN the age of big data, high-dimensional data have become commonplace in almost any application domain, and an abundance of methods and paradigms to explore such data have been devised. One of these is a framework we recently proposed, called the *Subspace Voyager* [35]. The Subspace Voyager aids in the exploration of a high-D data space by decomposing it into a continuum of generalized 3D subspaces which are visualized as scatterplot projections. Analysts can then explore these 3D subspaces individually via the familiar trackball interface while using additional facilities to smoothly transition to adjacent subspaces for expanded space comprehension. To make the selection of subspaces easier, the framework provides a set of data-driven subspace selection and navigation tools which guide users to interesting subspaces and views. The fluidity of the interface does not require users to ever think of data in their native high-D context. Rather, they can just go from one generalized 3D subspace to the next in a goal-directed manner, playfully using the interface elements designed for this purpose. The paper [35] describes all of these interactions in great detail and a video is also available on YouTube

In our user studies, when testing the utility and effectiveness of the Subspace Voyager, we observed that users frequently had trouble in visually distinguishing different clusters and also in locating outliers. While the study participants were generally content with the dynamic interface, some said that the data presentation in form of points, eventually colored by cluster ID, diminished their ability to get a sense of cluster overlap and their inter-relationships. While the motion parallax afforded by the trackball interface

- B. Wang was with the Visual Analytics and Imaging Lab at the Computer Science Department, Stony Brook University, NY. She is now with Google LLC. E-mail: icewang0317@gmail.com
- K. Mueller is with the Visual Analytics and Imaging Lab at the Computer Science Department, Stony Brook University, NY. E-mail: mueller@cs.stonybrook.edu

helped a great deal, it was nevertheless a transient effect which required continuous interaction for replay. This motivated us to develop the extension to the Subspace Voyager framework that is subject of this paper.

The method we propose seeks to recreate the transient experience with a reconstruction of a static shape from the point cloud. It abstracts cluster objects into geometric shapes and makes use of shading and occlusion effects to communicate the relationships previously only discernable via the dynamic display. All facilities of the Subspace Voyager are still supported in the enhanced framework, but now the trackball is used to inspect a set of 3D subspace shapes from different viewpoints and to transition to shapes located in adjacent 3D subspaces. Alternately, users can also use the point display as usual and only switch to the shape display occasionally to clarify certain relationships, using what we call a 'clarify' button.

Motivating our work is the fact that humans are highly adept at recognizing even subtle variations in shape and are able to do so at pre-attentive rates [18]. The framework we propose aims to take full advantage of these capabilities, in order to aid analysts in gaining a better understanding of their high-D datasets.

Interesting to our mission is also a recent study conducted in the Smithsonian Museum in Washington DC [1] where visitors were asked to rate certain abstract 3D shapes for aesthetics. The study suggested that curved 3D shapes are more attractive to humans, as opposed to non-curved ones, offering them more aesthetic pleasure. In fact, the 3D shapes our renderer produces are strikingly similar to these curved museum pieces, although they are shaped by data and not by artists.

Our paper is structured as follows. Section 2 provides a discussion on the proper use of 3D graphics in the display of non-spatial data. Section 3 presents related work. Section 4 provides theoretical background on the perception of shape and it also discusses the existing Subspace Voyager upon which this work is built. Section 5 describes our shape generation and rendering framework. Section 6 presents a set of case studies. Section 7 describes a user study and its results. Finally, Section 8 offers conclusions.

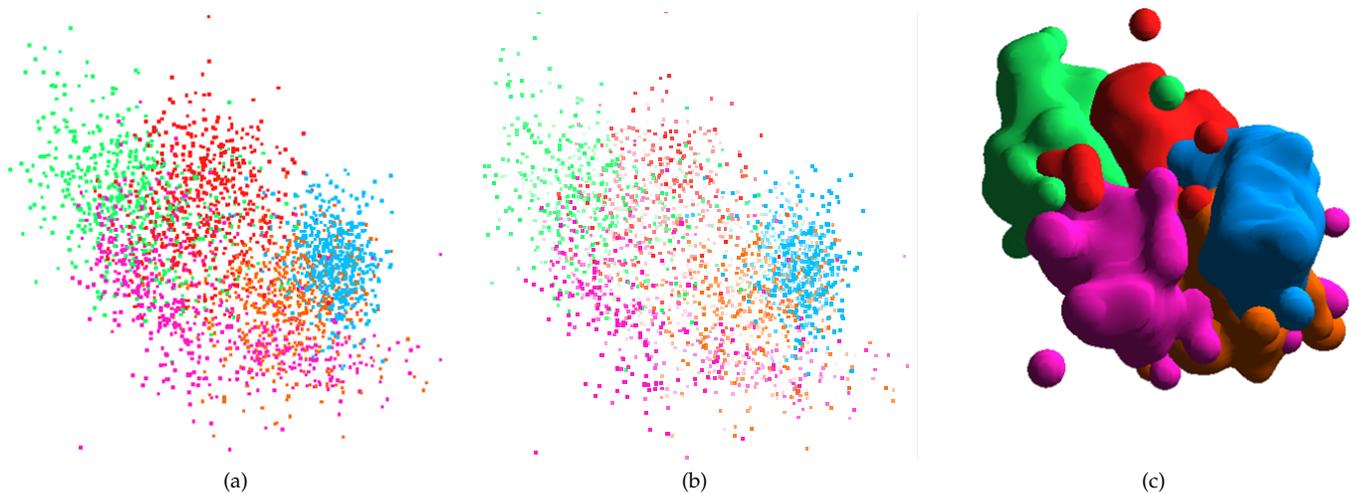

Fig. 1: High-dimensional point cloud projected into a generalized 3D subspace and visualized with the Subspace Voyager at an orientation selected using the trackball. The points have been divided into five clusters tagged by color. (a) Simple scatterplot projection; (b) with opacity depth cueing – points further back are rendered more transparent; (c) shaded display after converting the point cloud clusters into a set of solid models using the framework described in this paper.

## 2 USING 3D GRAPHICS IN NON-SPATIAL INFORMATION DISPLAYS: A DISCUSSION

Adding the third dimension for the display of data that do not have a direct geospatial reference is not inherently new. In fact, it has been a somewhat controversial topic over the past two decades. Robertson and Card, in collaboration with various co-authors, published numerous papers in the 1990s on this subject, extending planar trees to cone-trees [23], compiling web pages into 3D web books [2], and developing a 3D version of the 2D Data Mountain – a collection of possibly overlapping document thumbnails [22]. These representations were soon diligently evaluated by Cockburn and McKenzie in the early 2000s [4], [5], [6], [7]. Their essential findings were that while the 3D representations were often found more attractive than their 2D counterparts, they did not increase reader performance, and sometimes even reduced it.

Similar is true for 3D bar and pie charts. They are part of the repertoire of almost any plotting program and have been used in countless business graphics. Siegrist [26] showed that especially the perspective angle of the popular 3D pie chart has a significant effect on the accuracy at which viewers can estimate the graphed quantities. The same author also found that 3D bar charts require more time for evaluation, especially smaller bars.

Tory and colleagues [32], [33] confirmed these observations when testing a specific class of data spatializations [9] – the 3D landscape. The 3D landscape is a continuous terrain plot created from a 2D scatterplot, mapping height to either point density, a special attribute, or a classification tag [3], [37]. They found that landscape representations neither supported better memorability nor did they lead to better estimation speed or accuracy.

Many of the application mentioned above, such as cone-trees and 3D Data Mountains, used the third dimension as a handy way for resolving overdraw – by spreading items into the depth dimension. An additional aim was to make the displays more appealing to users, without any true purpose. This is especially the case for 3D bar and pie charts. The 3D landscape displays, on the other hand, in particular those tested by Tory and colleagues, used the third dimension for redundant attribute encoding [17] but in a way that was not overly helpful or successful.

Our application of extending an information display to 3D is fundamentally different from the above. We use it to better communicate cluster extent and shape, relations among clusters, and the significance of outliers. In that regard our display is more closely related to 3D modeling and CAD displays where the third dimension provides an additional level of object understanding, often supporting a triad of orthographic projections. Relevant are also the positive findings by Ware and Franck [36] who studied the added impact afforded by 3D renderings for distance estimation tasks, e.g., to assess paths in a complex network.

To illustrate our contribution, we provide Figure 1. Here, panel (a) shows an ordinary scatterplot projection of a generalized 3D subspace obtained using the Subspace Voyager. Panel (b) shows the same scatterplot projection but with points further away rendered less opaquely, mimicking fog in the scene. This allows for a slightly better appreciation of depth relationships. Finally, panel (c) shows a 3D rendering of the same set of points generated using the framework described in this paper. The cluster relationships are now clearly conveyed and could be clarified further by rotating it via the trackball or transitioning to an adjacent subspace.

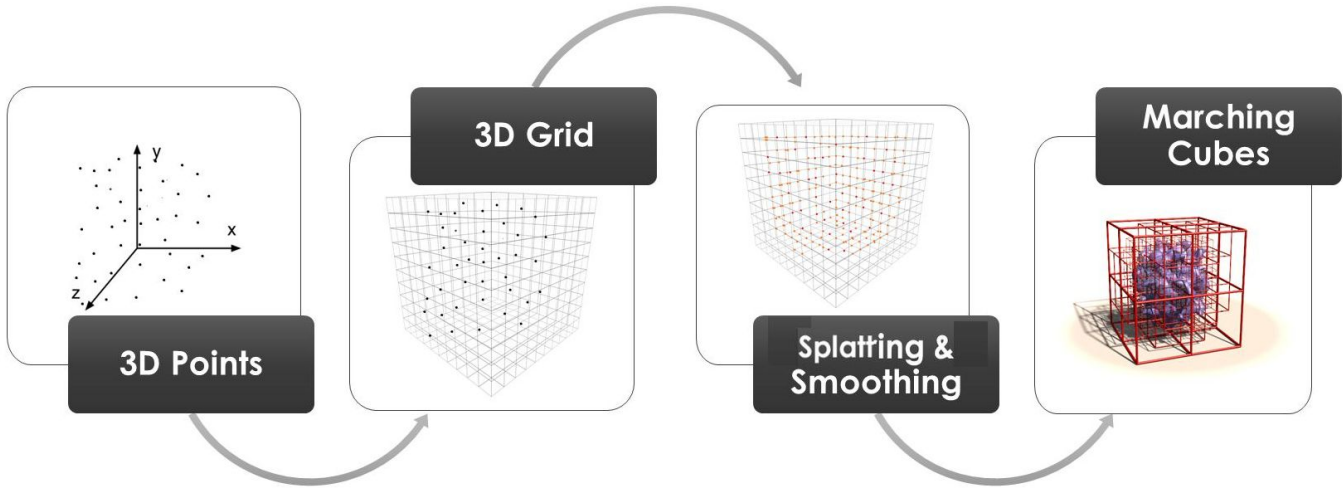

Fig. 2: Pipeline to construct a subspace shape from a 3D point cloud

## 3 RELATED WORK

One of the most widely used information visualization techniques for high-D or multivariate data is the scatterplot. The scatterplot projects the data onto two orthonormal vectors which gives clear insight into the trend of the data and the relation between the data dimensions associated with these vectors. The scatterplot matrix (SPLOM) [10] groups all axis aligned scatterplots arising from the multivariate data into a display matrix which enables users to inspect all pairwise relations at the same time. A caveat that both bivariate scatterplots and SPLOMs share is that they are all based on axis-aligned views. However, sometimes important phenomena may only be spotted in non-axis aligned projections. The Subspace Voyager (SV) [35] allows users to observe these non-axis aligned configurations.

Other methods that have exploited dynamic transitioning of scatterplots include ScatterDice [8] which restricts the transitions to motions between two SPLOM tiles, giving rise to a dynamic 3D point cloud projection display. Similar to our SV, the GGobi system [30] also employs trackball controls but it does not have the advanced subspace exploration facilities the SV's trackball interface provides. For example, with GGobi users cannot explicitly travel between adjacent subspaces and navigate the space via a map. Noteworthy in that context is also the work by Piringer et al. [19] who added 2D and 3D histograms on the bounding cube surfaces of a 3D scatterplot display to better convey point densities.

Sedlmair et al. [25] investigated the relationship of dimension reduction and visualization paradigm (2D and 3D scatterplots and SPLOMs) with regards to the ability of users to discern cluster separability. While they find that 3D scatterplots do not provide additional benefits for the particular task they studied, they also argue that 3D displays might be a good choice if the intent was to recognize cluster shapes. The original SV and also our present work both have this intent – their 3D displays do not only allow users to visualize static 3D scatterplots, but rather they let users interact with them to mentally reconstruct shape relations, first via dynamic points clouds and now as explicit shapes.

Several studies [13], [34] have shown that subspace analysis can decompose fairly high-dimensional datasets into a manageable set of subspaces of much lower dimensionality. The significant reduction that can be obtained is quite impressive – often these subspaces have less than five principal (non-axis aligned) dimensions. And so, while it might be infeasible to explore a native high-dimensional space with sequential 3D tools like the SV, once the space has been decomposed into a set of smaller independent subspaces such tools can be very powerful and meaningful to use.

The idea of representing high-D data with a surface-based representation has been pursued before. Sprenger et al. [27] and Rohrer et al. [24] both use implicit surfaces. The two approaches, however, differ in how the surface is constructed. The H-BLOB algorithm by Sprenger et al. first conducts a hierarchical cluster analysis and then fits an ellipsoid to each cluster's first three principal components. To construct the shape they tessellate the ellipsoid using Marching Cubes [14]. The framework by Rohrer et al., on the other hand, creates a star-shaped coordinate system in 3D space to represent the 14 most dominant principal components of the data space and constructs a blobby model [1] from it. Both methods provide visualizations of hierarchies using semi-transparent renderings of encapsulating hulls. A more recent contribution is that of Poco et al. [20] who first create an optimized point embedding of the high-D data into 3D and then construct a surface for each cluster using a variety of techniques – convex hull, blobs, and several surfaces derived from a 3D discrete Voronoi diagram. The implicit surfaces that arise from blobs and elliptical fits give rise to fairly roundish shapes without much surface detail while convex hulls create surfaces that have a rough faceted look. In contrast, our shapes are directly derived from the point cloud's density field yielding intricate surfaces that are a tight yet smooth "shrink-wrap" around the point cloud.

Finally, Mayorga and Gleicher presented Splatterplots [15] which are not shaded 3D surfaces in a technical sense but share some of their visual appearance characteristics. They are essentially colored overlapping regions of abstracted projected clusters that refine detail upon zoom. As zooming continues more points appear, but the density of points does not change.

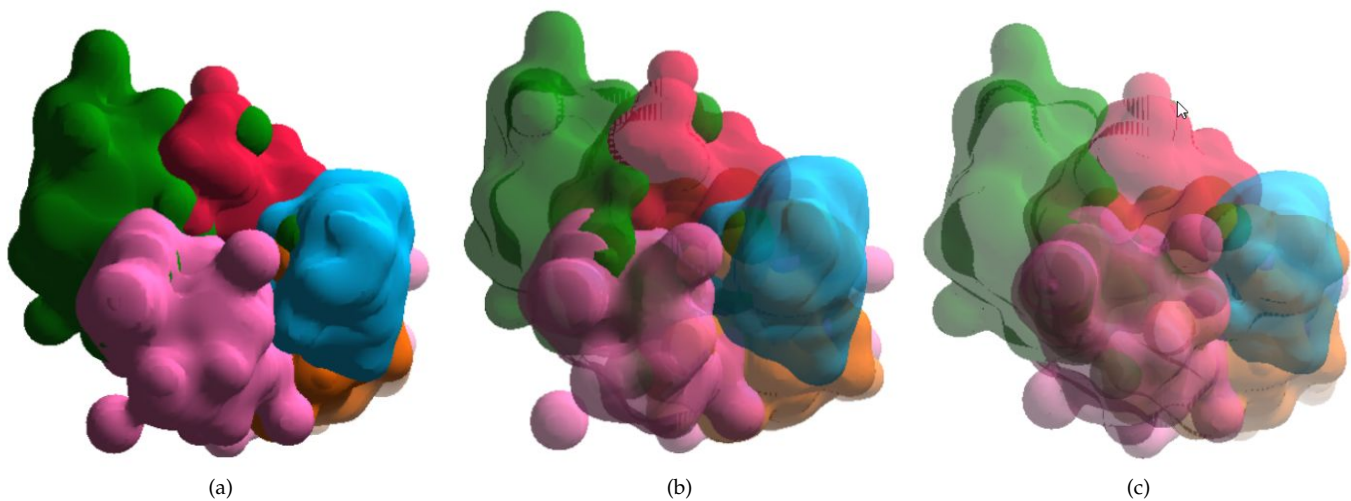

Fig. 3: Subspace shapes at different levels of opacity. (a-c): opacity = 1.0, 0.7, and 0.5.

## 4 SHADING FOR 3D SHAPE PERCEPTION

The shape-based paradigm we propose capitalizes on the human visual system's capability to rapidly derive 3D shape information of an object by assessing the variation of image intensities on the object's surfaces [21]. The intensity variations originate in the local differences of the amount of light reflected from the surfaces toward the observer's eyes. In the literature, these illumination-based appearance variations are referred to as *shading* and the ability to extract shape information from them is known as *Shape from Shading* [11].

The visual recovery of shape is synonymous to cognitively estimating a depth map. When it comes to the effectiveness of shading for enabling such estimations it has been shown [12] that basing a local surface's luminance on its normal is inferior to basing it on the amount of light impending on it. Essentially, the former applies to the sunny day scenario, while the latter refers to a cloudy day [12]. The cloudy day scenario embodies a much greater amount of diffuse lighting in which light can reach a surface element after bouncing off of one more intermediate reflectors.

*Ambient occlusion shading (AOS)* [16], [28] is an effective method to estimate this diffuse lighting. While it is substantially less accurate than the far more computationally complex global illumination paradigm, it can effectively darken small crevices in surfaces and so greatly improve their 3D appearance. AOS works by omnidirectionally probing the vicinity of a surface point on how much of it is occluded. As such, it elucidates the local geometries that are found in its visibility field, defined as the free space above the surface. AOS has become very popular for the visualization of molecular models, where it is able to clarify important features of biomolecular complexes, like pores, pockets, and cavities, which direct lighting cannot convey [29], [31].

## 5 SHAPE CONSTRUCTION AND RENDERING

Figure 2 outlines the shape construction process. Since we mostly deal with non-spatial data, these 3D points are irregularly sampled and are not perfectly aligned with the vertices of a volumetric grid. We solve this problem by splatting the points into a regular grid using reversed trilinear interpolation. After this step, we apply a 3D box filter on the grid to smooth out the points since otherwise the final shapes typically appear jagged. We use several rounds of smoothing. A slider allows users to balance the grid resolution and filter size as well as the number of iterations to obtain a shape that is sufficiently smooth but still preserves detail (see Figure 1c).

The shapes themselves are represented as closed meshes of polygons. We use the Marching Cubes algorithms [14] to convert the density values stored in the 3D grid to the polygon meshes – the last step of the pipeline shown in Figure 2. We can then set the opacities of the polygonal shapes to different values to better express overlap relationships. Figure 3) presents examples for different opacity levels.

Often clusters have dense cores with less dense (skewed) regions around them and outliers in the periphery. By using 3D iso-contouring and assigning smaller opacities to the outer contour polygon meshes we can produce displays like those shown in Figure 4). The transparency and the number of layers of the shapes can be controlled by adjusting the Opacity and Layers sliders in the control panel. Likewise, users have also the option to display detected outliers along with these semi-transparent shapes (see Figure 4b)

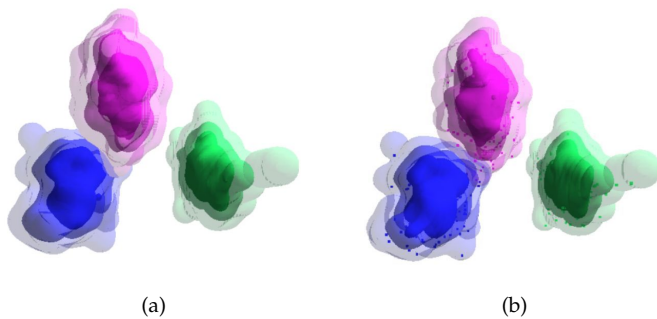

Fig. 4: : Multi-layer rendering mode (a) without and (b) with outliers

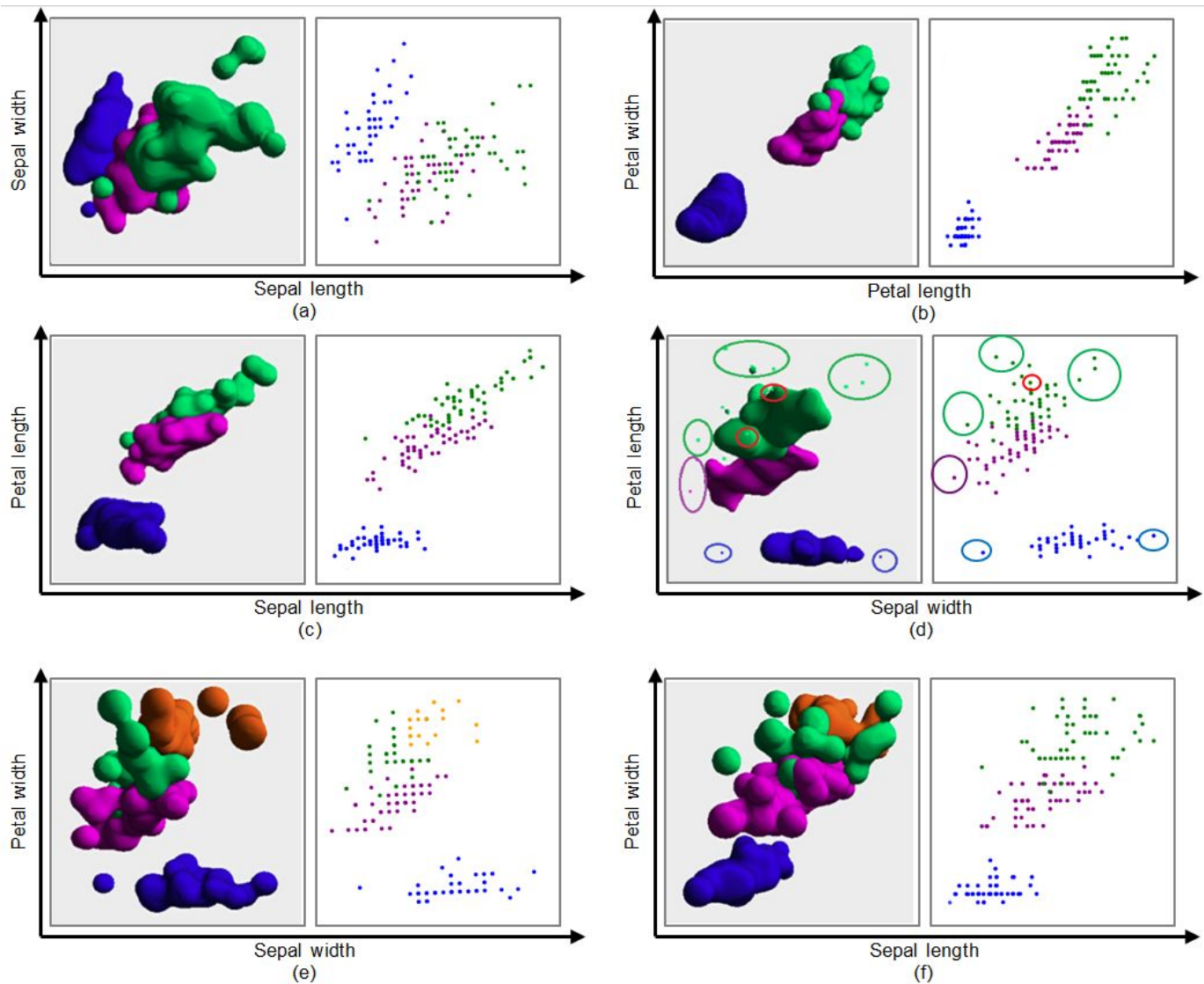

Fig. 5: Comparisons between scatterplots and our shaded shape representation (SSR) using the Iris dataset and different combinations of the attributes sepal length, sepal width, petal length, petal width, class label (a) The green and purple clusters are not separable in the scatterplot but our SSR suggests the possibility to use a plane parallel to the screen to separate them. (b)(c) From the scatterplots, only the 2D distribution of the data can be seen but our SSR allows users to observe the trend along the third dimension. (d) Outlier detection. Our SSR is able to detect outliers that are far away from the majority of points along the third dimension (red circle). (e)(f) Brushing and linking. A portion of the green cluster is painted yellow in (e), all points belonging to that portion form a new cluster for shape generation in (f).

Users can freely rotate and transition to adjacent subspace shapes within the trackball interface. While a new shape needs be generated upon subspace transitions, given reasonable data complexity this takes just a few seconds. We cache the previous subspace shape and so users can fluidly return to the previous shape and assess the high-dimensional structure of the multivariate point cloud.

## 6 THE CLARIFY BUTTON - EVALUATION

The shaded shape representation (call it SSR) has the unique capability to reveal cluster overlaps, proximity of outliers, and etc. which may be hard to see in the 2D scatterplots. For this purpose we have invented the magic "Clarify" button which, when hit, adds a third data (or other orthogonal) dimension to the view and renders a 3D shaded shape.

We use Fisher's Iris Flower data set [2] as an example to illustrate this feature. The Iris Flower data set has three classes with 50 instances each. Each class belongs to a different type of iris. The dataset has five attributes sepal length, sepal width, petal length, petal width, and the class label. Data analysts often use a scatterplot matrix (SPLOM) [10] to explore multivariate datasets of this nature. We use the first four attributes to generate half of the scatterplots in the SPLOM and place their corresponding SSRs to their left (see Figure 5). Essentially, each view on the left could be the 3D rendering that would result upon hitting the aforementioned Clarify button hen only the 2D scatterplot on the right is seen.

2. https://archive.ics.uci.edu/ml/datasets/iris

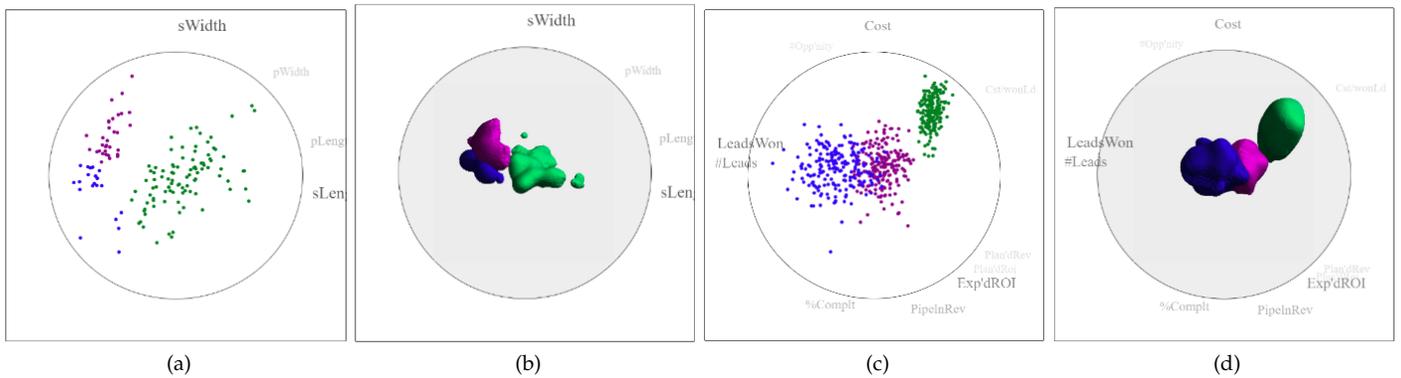

Fig. 6: Displays generated in the Subspace Voyager trackball for our user study. (a) standard multivariate scatterplot display for the Iris dataset; (b) shaded shape representation (SSR) for the same configuration as in (a); (c)(d) same paring as (a)(b) now for the Sales Campaign dataset.

### 6.1 Clarify separability

Let us assume a plant biologist, Tim, views the scatterplot display in Figure 5a on the right which is a rather complex one. While he can easily discern that the blue cluster of flowers is separate, he cannot tell for sure for the green and purple clusters. So he hits the Clarify button which produces the 3D view to the left in Figure 5a. He can now clearly see that the green cluster is in front of the purple one and a plane parallel to the screen seems to be able to separate them. This gives Tim good guidance on how to find a proper view where all three clusters can be separated. Since the current view is Sepal width over Sepal length, he now knows that the other attributes, Petal width and Petal length, or both, are also involved in the classification.

### 6.2 Clarify outliers

Tim moves on and inspects a 2D scatterplot of Sepal width and Petal length (Figure 5d, right). Looking at this display he suspects that there might be some flowers that are different from the main stream – outliers. To get more insight he hits the magic Clarify button and the system produces the 3D display on the left. Now he clearly sees that the points he suspected to be outliers (circled in green and blue) are indeed outliers. But he also discovers an outlier he did not suspect – the one circled in red. This point is indistinguishable from the others in the 2D scatterplot, but pops out easily in the associated 3D SSR. While Tim could have possible found this rare flower in a different 2D scatterplot, the 3D SSR is more direct and does not require a context switch to a different set of dimensions. Rather, it uses the same dimensions than the 2D scatterplot, just using shading to disambiguate it.

### 6.3 Engage and explore further

Next, Tim calls up the 2D scatterplot that visualizes Petal length vs. Petal width ((Figure 5(b, right). He sees three lean shapes that spread along the diagonal, suggesting a strong correlation for the green and purple classes. The blue class, on the other hand, does not seem to have a trend. Tim thinks that this is interesting but he is not certain what 2D scatterplot to look at next. He presses the Clarify button (producing the view on the left) and sees that all three classes seems to protrude into the depth direction, almost inviting him to follow to see what this trend is all about. He knows that this third dimension has to be one of the remaining ones, or a combination of them. So he gets curious again and exchanges Petal with by Sepal length (see (Figure 5c) to learn more about his data, and so on.

### 6.4 Brushing and linking in 3D

Brushing and linking are important activities in interactive cluster analysis. Tim decides that he would like to break part of the green class into a separate class since he thinks that it an important subset of flowers. Tim prefers to do this in 3D since he used to painting 3D objects. He picks the yellow color and brushes on a portion of the green surface. Our system automatically assigns the tagged surface points, as well as points underneath, to a different cluster. This produces Figure 5e, left. Next, Tim uses the trackball interface to go to a different 3D subspace (Figure 5f, left) where new shapes are generated according to the new axis dimensions. Tim observes that the tagged yellow cluster is now rotated to the back. This gives Tim an overall idea of the dynamics of what he thinks are the interesting parts of the data.

## 7 USER STUDY

To evaluate the pros and cons of the SSR we conducted a user study with nine graduate students - eight males and one female of diverse cultural background (two North American, one European and six Asians). We presented the same data in both the 2D scatterplot representation and the 3D SSR within our Subspace Voyager framework and asked the participants to accomplish various data analytics tasks.

We began with individual training sessions where each participant heard the following message: "Hi, we have previously built a 2D scatterplot-based visual analytics framework where the 2D data points are displayed inside a virtual trackball, like the trackball often used in video games. It acts on a model of the 3D world and has an imaginary third axis going into the plane. Now, in a new version of this interface we have added a 3D shape display to this virtual trackball and we would ask you to test the pros and cons for both of these data representations".

Next, we showed the 2D view and the 3D view for one projection of the Iris dataset side by side (Figure 6a,b). We said, "This is an example, the Iris Flower dataset. There are three different kinds of iris flowers and they are represented in three different colors. These data capture the different petal length, petal width, sepal length and sepal width for these irises. Our trackball displays the dimensions along a virtual circle. The font size and the opacity of a particular dimension indicates the influence of this dimension for the projection shown. For example, in this projection, sepal width sepal length are the dominant dimensions. The green iris type has longer sepals than the other two." We gave the participants time for questions and asked for consent for video recording. None of the participants was color blind.

### 7.1 Tasks and procedure

The three tasks were designed to compare the two representations from both static and dynamic perspectives. We used the sales campaign dataset for all of them and gave all participants a brief introduction of it. This dataset consists of 900 data points (one per sales person) and 10 attributes: %Completed, #Leads, Leads Won, #Opportunities, Pipeline Revenue, Expected ROI, Actual Cost, Cost/WonLead, Planned Revenue, and Planned ROI. There are three pre-clustered sales teams.

For the first task, we wanted to test if the participants could reach similar conclusions from both a fixed 2D scatterplot display and a fixed 3D SSR display. We showed the participants two screenshots of the same projection of the dataset, one in 2D scatterplot representation, one in SSR (Figure 6c,d). We showed odd numbered participants the 2D representation first while even numbered participants saw the SSD first to minimize learning effects. We asked all participants the same question – "What can you conclude from this fixed projection?" – for both representations.

For the second task we wanted to test if the participants could interact with both representations and if they had any preference. We loaded our Subspace Voyager and told them: "The trackball in our system supports many different interactions. One of which is 3D rotation. Pressing the left mouse down while moving the mouse will rotate both the 2D and the 3D representations. By checking and unchecking the Shape checkbox, you can switch between those two representations. [3]" We let the participants play with the system for as long as they wanted to and then asked them the following question: "How do you like the 2D scatterplot representation / 3D shaded shape representation when combined with rotation?"

The third task was to let the participants analyze the sales campaign dataset freely. Specifically, we asked them to uncover the different sales strategies the three different teams chose. For this task, we wanted to analyze the time the participants spent on each representation. After the three tasks, we asked all participants for their opinions with respect the two different representations.

In the end, we showed the participants the campaign data in the transparent shaded shape representation with embedded points (Figure 4b) and solicited their thoughts.

---

3. We had relabeled this button from "Clarify" to "Shape" as to not bias the subjects toward the goal of our study.

### 7.2 Results

#### 7.2.1 Task 1: fixed projection

For this task, most participants came to very similar conclusions for both the 2D and the 3D representations. They said, e.g.: "The green team is very different from the other two", "Leads Won, Expected ROI and Cost are well expressed in this view", "The green team has higher Cost" and "The blue team generates more Leads and wins more Leads".

Five participants noticed the depth cue that is only expressed in the 3D representation: "The green team seems to be sticking out" and "In the points view, I noticed that that blue and purple is overlapping in one surface, but in 3D I actually can tell that blue is in front of purple".

#### 7.2.2 Task 2: trackball rotation interaction

All participants mastered the rotation interaction for both representations easily and could effortlessly switch between them. When asked about their opinions on the two representations when combined with rotation, eight out of nine participants expressed that they preferred the SSR when rotating. The most dominant reasons were: "Rotating a shape is more natural while rotating points is a bit messy.", "Shapes are better at defining boundaries and show clusters separations much more clearly." and "Shapes get rid of outliers and allow me to focus on the main structure."

#### 7.2.3 Task 3: Data analysis

All participants managed to come to some conclusions about the different sales strategies the three sales teams used. Since the default representation is the Subspace Voyager in the 2D scatterplot view, all participants started the analysis from there. After a while, eight participants switched to 3D. Six of nine participants spent significantly more time using SSR when analyzing the data. Two participants spent more time using 2D scatterplots while one participant only used 2D scatterplot. One of the two participants who relied mostly on 2D scatterplot mentioned later that he just went with the default 2D view and he would have used the 3D view more if it had been the default.

#### 7.2.4 Conclusions and insights

The dominant feedbacks we received on both representations were: "I like 3D with rotation a lot.", "3D provides better separability for clusters and gives a much clearer overall understanding of the data." and "3D hides outliers and let users focus on the main data." One participant mentioned that he would use 3D as an aid to 2D in the sense that "for analyzing overall relationships, 3D is better but for individual points, 2D is better." Another person mentioned that he liked both and "its easiest to see everything in 3D but seeing all data is also important." However, some participants also expressed a concern: "3D hides outliers and I think 2D is still better for point level analysis."

Then we showed the participants the combined representation with transparent shaded shapes enclosing all the points (Figure 4b). Six participants really liked it and said things such as "This is perfect", "This allows me to see the overall information with all points", "It allows me to see inside the shapes", "The outliers are so clear now" and "it helps me to understand how the shapes are built."

Three participants said it was a useful view but had too much information and they still preferred to have the two representations separately.

We can conclude from the user study that participants prefer SSD for an overall understanding of data, consider it a much clearer presentation with better separability than scatterplot, and see it as a less distractive representation which allows users to focus on the main data. The biggest concern is that SSR hides information such as outliers which can be eliminated using the combo representation.

## 8 CONCLUSION

The use of 3D rendering in information visualization has been a controversial topic. We believe we have presented a compelling use for it, namely its ability to convey the shape of multivariate point clouds. Shape recognition is an innate quality of the human visual system and we found that users find our 3D displays appealing and insightful. In our user study, many chose to spend more time with the 3D shape display then with the matching dynamic scatter plot display.

## ACKNOWLEDGMENTS

This research was supported by NSF grant IIS 1527200 and the MSIP Korea "ITCCP Program" directed by NIPA.